**PAPER • OPEN ACCESS**

# Key-Point Interpolation: A Sparse Data Interpolation Algorithm based on B-splines

View the article online for updates and enhancements.

## You may also like

- Interpolation-based outlier detection for sparse, high dimensional data
  Wanghu Chen, Zhen Tian and LiZhi Zhang

- Characterisation of freeform, structured surfaces in T-spline spaces and its applications
  Jian Wang, Renqi Zou, Bianca Maria Colosimo et al.

- Research and Development of Sampled-data Interpolation Algorithm Software in CNC System Based on the Visual C++
  Linguo Yang and Guozhen Zhang

# Key-Point Interpolation: A Sparse Data Interpolation Algorithm based on B-splines

**Bolun Wang[1,2], Xin Jiang[1,2,3] *, Guanying Huo[1,2,3], Cheng Su[1,2,3], Dongming Yan[4], and Zhiming Zheng[1,2]**

[1]Key Laboratory of Mathematics, Informatics and Behavioral Semantics (LMIB), School of Mathematical Science, Beihang University, Beijing, China
[2]Peng Cheng Laboratory, Shenzhen, Guangdong, China
[3]Zhengzhou Aerotropolis Institute of Artificial Intelligence, Zhengzhou, Henan, China
[4]National Laboratory of Pattern Recognition, Institute of Automation, Chinese Academy of Sciences, Beijing, China

*Corresponding author email: 09436@buaa.edu.cn

**Abstract.** B-splines are widely used in the fields of reverse engineering and computer-aided design, due to their superior properties. Traditional B-spline surface interpolation algorithms usually assume regularity of the data distribution. In this paper, we introduce a novel B-spline surface interpolation algorithm: KPI, which can interpolate sparsely and non-uniformly distributed data points. As a two-stage algorithm, our method generates the dataset out of the sparse data using Kriging, and uses the proposed KPI (Key-Point Interpolation) method to generate the control points. Our algorithm can be extended to higher dimensional data interpolation, such as reconstructing dynamic surfaces. We apply the method to interpolating the temperature of Shanxi Province. The generated dynamic surface accurately interpolates the temperature data provided by the weather stations, and the preserved dynamic characteristics can be useful for meteorology studies.

## 1. Introduction

B-splines are widely used in the fields of reverse engineering and computer-aided design, due to its superior properties such as global smoothness, local support, etc. [1][2][3][17]. Traditional B-spline surface interpolation algorithms usually assume that the data points are distributed in grids or in rows. However, the majority of actual measured data are naturally sparse and non-uniform, such as the data obtained from LIDAR scanners and weather stations [6]. To interpolate sparse data points, traditional algorithms usually introduce degrees of freedom, and require a large number of control points, which may lead to longer runtime and larger memory usage.

In this paper, we introduce a novel B-spline surface interpolation algorithm: KPI, which interpolates sparsely and non-uniformly distributed data points. Our algorithm is a two-stage method [11]. Firstly, we generate a gridded dataset to meet the requirements of B-spline interpolation. Then, we use our Key-Point Interpolation method to generate the control points. Our algorithm can be extended to higher dimensions by properly assigning the key points, thus can be used to construct dynamic surfaces [3][5][7]. We use our method to interpolate the temperature of Shanxi Province. The result shows that our method generates a dynamic surface which interpolates the sparse temperature data accurately, and the dynamic characteristics preserved in the 3 dimensional B-splines can be useful for meteorology studies.

## 2. Related Work

Previous works usually assume that the data points are distributed in grids or in rows. [14] and [2] use averaging method to calculate the knot vectors from the data points. If the data points are sparsely distributed, these methods will create a large number of control points. [7] designs an algorithm to interpolate sparse data points. In each iteration, [7] doubles the density of the knots, until the data points are interpolated accurately, but the number of required control points highly depends on the distribution of the data points. [13] and [15] use lofting method to obtain the interpolation surfaces. In [13], a sufficient condition for the existence of the interpolation surface is introduced. By employing a flexible knot insertion method, [13] can reduce the control points effectively. [15] further improved the surface smoothness by dividing the curves into groups. Lofting methods can be efficient when interpolating large number of data points, but they usually require that the data points are sampled in rows.

Kriging [10][16] is a standard method to predicate space structure and randomicity of parameters from sparsely sampled data points in geostatistics, by performing optimal linear unbiased interpolation estimates. We use Kriging to construct a gridded dataset out of the input sparse data, thus it can meet the requirements of our KPI control points solving procedure.

## 3. KPI Method

Let $Q = \{q_0, q_1, \cdots, q_{m_1}\}$ be a point set in $\mathbb{R}^d$, and $\tilde{Q} = \{\tilde{q}_0, \tilde{q}_1, \cdots, \tilde{q}_{m_2}\}$ be a subset of $Q$. Let $f_1$ be a function on the set $Q$, and $f_2$ be a function on $\tilde{Q}$. Suppose the valve of $f_2$ on $\tilde{Q}$ is given. $f_1(\tilde{q}_i) = f_2(\tilde{q}_i), \forall \tilde{q}_i \in \tilde{Q}$. The purpose of the interpolation is to find the function $f_1$ on $Q$ with the given value on $\tilde{Q}$. In this paper, a 3 dimensional B-spline function $M$ which reconstructs dynamic surfaces is the function $f_1$ over $Q$, while the data points $\tilde{Q}$ is given as the form of sparsely sampled data. Our KPI method is introduced as the following 2 steps:

### 3.1. Dataset Generation

In this step, we firstly parametrize the dataset, then generate a gridded dataset using Kriging. Constrained Delaunay Triangulation (CDT) method [12] is employed to generate a triangle mesh for each surface. However, if the border of each surface is over complicated (for example, when interpolating the temperature or the topographic information bounded by the border of a certain region), the number of border points could be a few orders of magnitude higher than the sparse data, which may lead to degenerated triangles. To avoid that, we simplify the border until the number of border points is comparable with the number of sparse data points. We use Dmitry and Zsolt's method [8] to detect feature points of the border, and use the feature points as the vertices of the simplified border. Then triangulate the sparse data $Q$ and feature points $\tilde{B}$ of the border $B$ to obtain the triangle mesh $M = (v, E, S)$ of each surface, where $v = \tilde{Q} \cup \tilde{B}$ represents the vertex sets and $E$ presents the edges of the triangulation mesh. We apply the method into a temperature data interpolation problem. Figure 1 is an example of border simplification and triangulation of the weather stations in Shanxi Province, China. In figure 1 (a), the red dots on the border are the selected feature points, (b) and (c) show the simplification result of the border. Figure 1 (d) shows the triangulation result.

As the triangulation $M$ is obtained, we use mean value coordinates method [9] to parametrize the data. Since the data points are randomly distributed, the parameters are extremely non-uniform and dense.

Here we provide a perturbation method to reduce the number of parameters, while maintaining the topology of the triangle mesh.

We use two parameters $C_1$ and $C_2$ to bound the perturbation in u and v direction, separately. The parameters for each vertex $v_i$ are denoted as $r_i = (u_i, v_i)$. In the parameter domain $M' = (r, E', S')$, for the u parameters $u_i \neq 0, u_{i+1} \neq 1$, if $|u_{i+1} - u_i| < C_1$, we merge the two u parameters by setting $u_{i+1} \leftarrow u_i$; For each $v_j \neq 0$, $v_{j+1} \neq 1$, if $|v_{j+1} - v_j| < C_2$, set $v_{j+1} \leftarrow v_j$. In this way, any two u (or v) parameters closer than $C_1$ (or $C_2$) will be merged. Iterate this procedure with increased $C_1$ and $C_2$ values, the number of parameters can be progressively reduced.

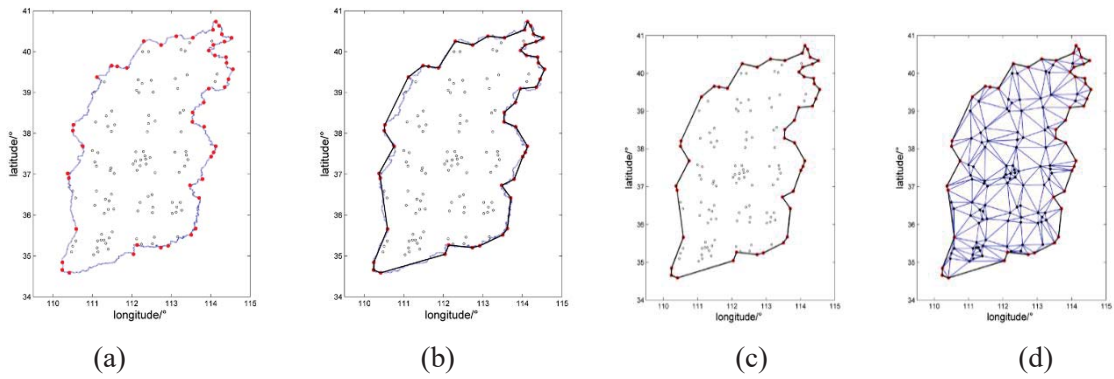

(a)　　　　　(b)　　　　　(c)　　　　　(d)

**Figure 1.** Border simplification and triangulation of the data points.

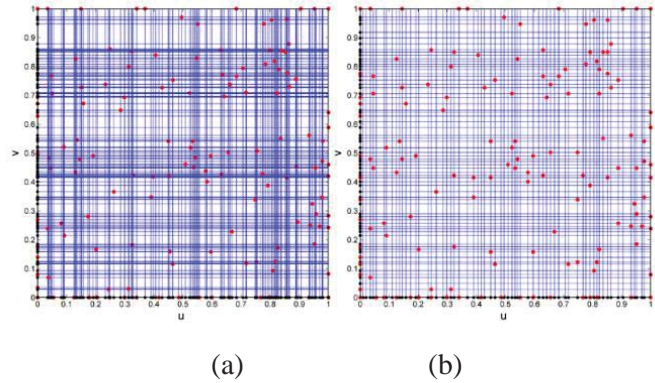

(a)　　　　　(b)

**Figure 2.** Original u and v parameters and parameter perturbation result.

To avoid parametrization failures (i.e., there is no one-one mapping between the old and new parametrization), we add an extra check when iteratively reducing the parameters. We assume that for a triangle $T_{i,j,k}$ defined by its three vertices $v_i, v_j$ and $v_k$ whose parameters are $r_i$, $r_j$ and $r_k$. After perturbing at least one of the 3 parameters of the vertices, the new parameters become $\tilde{r}_i, \tilde{r}_j$ and $\tilde{r}_k$. We check if the orientation of the 2D triangle in the parametric domain is the same as before, by checking if $sign\left(\det\begin{pmatrix} e_{ij} \\ e_{ik} \end{pmatrix}\right) = sign\left(\det\begin{pmatrix} \tilde{e}_{ij} \\ \tilde{e}_{ik} \end{pmatrix}\right)$, where $e_{pq} = r_q - r_p, (p, q \in \{i, j, k\})$, the function $sign(x)$ returns the sign (+1, -1, or 0) of the scalar value $x$, which is the determinate of a $2 \times 2$ matrix. After perturbation, if the orientation of any triangle doesn't hold, we withdraw the perturbation and continue to perturb another pair of u or v parameters. Here we choose $C_1 = C_2 = 0.5L$ as the initialized perturbation bound, where $L = \arg\min_{i,j=1\cdots m} |r_i, r_j|$.

Iterating under the orientation check, the number of parameters can be reduced progressively. Generally, within a few iterations, the number of the parameters can be reduced to a suitable quantity. The generated parameters are denoted as $\bar{U} = \{\bar{u}_0, \bar{u}_1, \cdots, \bar{u}_{m_1}\}$ and $\bar{V} = \{\bar{v}_0, \bar{v}_1, \cdots, \bar{v}_{m_2}\}$. In figure 2 (a), the original parameters are very dense and non-uniform. After perturbation (figure 2 (b)), the parameters are reduced, and become more uniform.

To ensure the smoothness of the interpolation surface, our method generates a gridded dataset using Kriging interpolation method [10]. The generated dataset $Q_{i,j}, i = 0, \cdots, m_1, j = 0, \cdots, m_2$ of each time step is composed of two parts: the original sparse data and the Kriging generated data.

### 3.2. Control Points Solving based on KPI

A 3 dimensional B-spline function [4] can be expressed as follows:

$$M(u,v,t) = \sum_{i=0}^{n}\sum_{j=0}^{m}\sum_{k=0}^{l} N_{i,p}(u) N_{j,q}(v) N_{k,r}(t) P_{i,j,k}, \quad (1)$$

Where $P_{i,j,k}$ are the control points, $N_{i,p}(u), N_{j,q}(v)$, and $N_{k,r}(t)$ are B-spline basis functions of degree $p, q$, and $r$. The knot vectors are $U$, $V$ and $T$. Solving the control points of the dynamic surface can be achieved by 3 levels of iterations of curve interpolations. In the first level, the data points to be fitted are the sparse data points and the Kriging generated points. In the second and the third level, the data points are the control points of the curves of the previous levels. We use the following equation to generate a KPI curve:

$$\min f = \sum_{i=0}^{m_1}(Q_{i_1} - C(\bar{u}_{i_1}))^2$$
$$s.t. Q_{i_2} = C(\bar{u}_{i_2}), i = 0,1,\cdots,m_2, \quad (2)$$

where $Q_{i_1}, i = 0,1,\cdots,m_1$ are the non-key points, $Q_{i_2}, i = 0,1,\cdots,m_2$ are the key points. $\bar{u}_{i_1}$ and $\bar{u}_{i_2}$ are the parameters of $Q_{i_1}$ and $Q_{i_2}$. The curve $C(u)$ interpolates the key points, and approximates the points generated by Kriging. It is crucial to determine the key points when iteratively constructing interpolation curves. The key points are determined as follows:

Step 1. For the first level, where the data points are the sparse points and the Kriging generated points, we directly set the sparse data points as the key points.

Step 2. For the second or the third level, the data points are the control points of the previous level. Without loss, we assume one key point in the previous level is $Q_{r,k}$, fitted by the curve $C_k(u) = \sum_{j=0}^{m} b_j N_{i,p}(u)$ and the parameter of $Q_{r,k}$ is $\bar{u}_r$. Then, according to the local support property of B-spline surface, there are at most $p+1$ non-zero B-spline bases $N_{i,p}(\bar{u}_r)$. Then, we set the $p+1$ corresponding control points $b_j$ ($N_{j,q}(\bar{u}_r) \neq 0$) as the key points of the next level.

There are many methods [2] to ensure the equality constraints in equation (2) can be satisfied. We iteratively inserting knots, until at most one key point whose parameter falls into each interval (e.g., interval $[u_i, u_{i+1}]$) of the knot vector of the curve. The generated knot vectors guarantee that the equality constraints can be satisfied.

## 4. Application

We interpolate the temperature of Shanxi Province, China, using our KPI algorithm. The data contains temperature data for each hour provided by 98 weather stations, from 1:00, October 15, 2017 to 24:00, October 17, 2017. Figure 1 shows the result of border simplification. The original border which is presented by 6102 points (shown in Figure 1(a)) is simplified into a polygon with 49 vertices, while maintaining the basic shape of the original border. The data was provided by China Meteorological Data Sharing Service System (http://data.cma.cn/en). To parameterize the data points, the 98 station points and 49 vertices of the simplified border are triangulated with Constrained Delaunay Triangulation method in a two-dimensional plane. The triangulation result is shown in Figure 1(b, c, d). Using mean value coordinates, the triangle mesh is parametrized into a quad parametrization domain shown in Figure 2(a). The number of u parameters and v parameters are 124 and 120, respectively. From Figure 2(a), we can see that the parameters are extremely dense and non-uniform. Process the parameters with our perturbation method. With 3 iterations, the numbers of u and v parameters are reduced to 66 and 68, respectively (shown in Figure 2(b)).

By Applying Kriging, the gridded data points $Q_{i,j,k}$ $(i=0,\cdots,65, j=0,\cdots,67, k=0,\cdots,71)$ are generated. A dynamic surface which contains $41\times 45\times 74$ control points is constructed (figure 3). The dynamic surface accurately interpolates the temperature data (with the maximum interpolation error $9.95\times 10^{-14}$). The surface is smooth since the Kriging generated data follows the trend of the sparse data. Figure 3 (a), Figure 3 (b) and Figure 3 (c) are the temperature interpolation results at 4:35, October 15, 2017, 14:00, October 15, 2017 and 10:15, October 16, 2017.

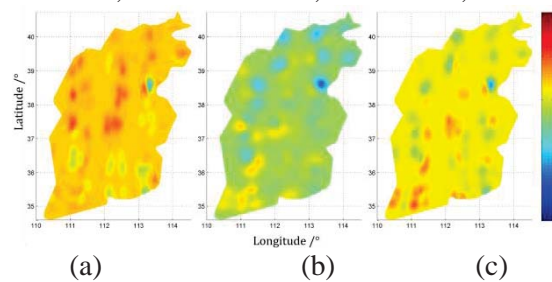

(a) (b) (c)

**Figure 3.** Temperature interpolation results of Shanxi Province with KPI method

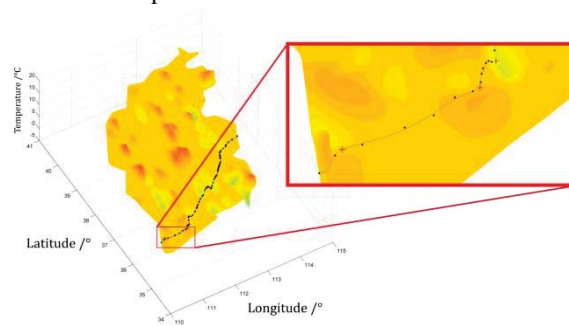

**Figure 4.** An iso-u curve which interpolates the sparse points (red crosses) and approximates the Kriging generated points (black dots).

Figure 4 shows an iso-u curve, where the red crosses are the original data provided by the weather stations and the black dots denote the data points generated by Kriging. It can be seen that our KPI method interpolates the original data points precisely, while approximates the Kriging generated data points.

## 5. Conclusion

In this paper, a KPI method for dynamic B-spline surface reconstruction is introduced. As a two-stage B-spline interpolation method, our method firstly constructs a gridded dataset, and then generates control points to interpolate the sparse data points. To construct the dataset, the sparsely distributed data points of each time step are triangulated. Then the meshes are parameterized in quad-domains using mean value coordinates method. We also design a parameter perturbation method which reduces the parameters to reduce the computation of control points solving. The control points are solved by iteratively assigning key points and solving curve KPI. Our method can generate a smooth dynamic surface which interpolates sparsely distributed data points given in a time sequence.

**Acknowledgments**
This work has been supported by the National Key Research and Development Program of China (Grant No.2020YFA0713700)) and the Natural Science Foundation of China (Grants No. 12001028, No. 61772523).